# User Detection and Response Patterns of Sycophantic Behavior in Conversational AI

KAZI NOSHIN, University of Illinois Urbana-Champaign, USA

SYED ISHTIAQUE AHMED, University of Toronto, Canada

SHARIFA SULTANA, University of Illinois Urbana-Champaign, USA

Despite growing attention to LLM sycophancy from researchers and developers, users' own experiences of this behavior remain underexplored. We examine how everyday users experience AI sycophancy through *Reddit* discussions. Using our ODR Framework which maps user experiences through observation, detection, and response stages, we find that users identify sycophantic behavior through methods like cross-platform comparison and consistency testing. They employ various mitigation strategies, including persona-based prompting and specific language engineering techniques. Our findings suggest that sycophancy does not have a uniformly negative effect; its impact differs by context. Users facing trauma, mental health struggles, or isolation often actively seek affirmative AI responses for emotional support. Users construct both technical and informal theories to explain sycophantic outputs. Users construct both technical and informal theories to explain sycophantic outputs. These findings suggest eliminating sycophancy entirely may be misguided. We argue for context-aware AI design that balances risks against benefits of affirmative interaction, with implications for user education and system transparency.

CCS Concepts: • **Human-centered computing → AI Sycophancy**; **Web-based interaction**; **Social networking sites**; **Reddit**.

Additional Key Words and Phrases: Ethics, Justice

## 1 Introduction

Widespread global usage of Large Language Models (LLMs) involves daily engagement from users [10, 53], across various tasks such as information retrieval [25, 57], emotional support [14, 22], creative writing [20, 56], coding assistance [27, 39], and decision-making [19, 23, 33]. However, concerns have emerged around sycophancy [46, 54], the tendency of conversational agents to align outputs with a user's perceived preferences, beliefs, or self-image, often compromising factual accuracy [9, 18, 42]. This behavior manifests in various forms [13, 46]: models may agree with incorrect statements, mirror user biases, provide overly positive feedback, or adjust responses based on perceived user expertise. Sycophancy can reinforce biases [24], spread misinformation [11], and undermine critical thinking [37], erode honest feedback, create echo chambers that strengthen pre-existing misconceptions, and damage trust in AI systems when users recognize the insincerity [51]. As reliance on AI grows for education [28] and decision-making, sycophancy poses risks to judgment quality [12]. Despite these concerns, little empirical work has examined how

Authors' Contact Information: Kazi Noshin, knoshin@illinois.edu, University of Illinois Urbana-Champaign, USA; Syed Ishtiaque Ahmed, ishtiaque@cs.toronto.edu, University of Toronto, Canada; Sharifa Sultana, sharifas@illinois.edu, University of Illinois Urbana-Champaign, USA.

users including both those with and without technical expertise, experience and respond to sycophantic behavior in real-world interactions. Understanding user perspectives is crucial for informing system design, user education, and determining whether technical definitions align with lived experiences and whether interventions should be tailored to specific user groups. To address this gap, we investigate AI sycophancy through the lens of user experiences obtained from Reddit, where individuals openly discuss their interactions with ChatGPT and other LLMs. By analyzing these views, we identify patterns in how users encounter and respond to sycophantic behavior across diverse contexts and use cases. Our research questions are:

*RQ1:* How does sycophancy prevail in ChatGPT's responses across different interaction contexts?
*RQ2:* How do users detect sycophantic behavior in AI chatbots during their interactions?
*RQ3:* How do users respond to the sycophantic behavior of AI after detection?

While addressing this questions from Reddit data analysis, we develop the Observation-Detection-Response framework (ODR, henceforth) of user experience that captures three critical dimensions of LLM sycophancy: how users define AI sycophancy, identify sycophantic behavior, and respond through emotional reactions (affective), mitigation strategies (behavioral), and explanatory sense-making. Our qualitative analysis reveals harmful, harmless, and even beneficial nature of LLM sycophancy. We identify user-developed detection and mitigation techniques, and examine how users explain sycophancy based on their knowledge and folk theories.

We offer three contributions to human-AI interaction research (HAI). **First**, we document user-developed sycophancy detection techniques beyond existing frameworks, including cross-platform comparison and inconsistency analysis. **Second**, we identify user-adopted mitigation strategies from persona-based prompts to specific language patterns such as positive and imperative tones. **Third**, we provide empirical evidence that sycophantic behavior serves therapeutic functions for users processing trauma, managing mental health challenges, or experiencing chronic isolation, revealing a complex dynamic where excessive agreeableness operates simultaneously as risk and resource. Our findings suggest context-aware AI design approaches that balance transparency and user education with the emotional support needs of vulnerable populations, rather than discarding sycophancy.

**Ethical Considerations Statement.** Reddit data is publicly available, so IRB approval was not sought for this analysis. However, we followed HCI community guidelines for protecting pseudonymous research participants [8, 36] and transparency in qualitative research [49]. All quotes were pseudonymized and manually paraphrased. We then searched for each paraphrased quote on Google to ensure its anonymity and that it could not be traced back.

## 2 Related Work

### 2.1 Human-AI Interaction and Sycophantic Behavior

Recent work in HAI has identified sycophancy as a critical challenge for trustworthy and responsible AI. LLM–based conversational agents often prioritize agreement and validation, over accuracy, raising concerns about user trust and decision-making [12, 48]. Experiments show sycophantic behavior reduces trust when paired with anthropomorphic cues such as high friendliness, as users perceive excessive agreement as inauthentic or manipulative [48]. AI systems agree with users far more than humans do, even validating harmful behaviors like deception or manipulation [12]. If users prefer this validation, it reduces users' willingness to repair relationships or act prosocially [12]. Technical surveys link sycophancy to hallucination and bias [35]. Uncritical validation may exacerbate harmful beliefs or behaviors, particularly in vulnerable populations [15], causing people to overestimate their own abilities [44], and enable fraud, political manipulation, and loss of human oversight [40]. Research suggests that sycophantic tendencies stem from alignment techniques. Reinforcement Learning from Human Feedback (RLHF) introduces bias where models prioritize agreement to maximize reward [5, 43]. Moreover, interaction context significantly amplifies sycophantic behavior [26]. Models also remain highly susceptible to sycophancy in addressing subjective opinions or user-provided hints [43].

### 2.2 Detection Techniques of Sycophantic Behavior

Detecting sycophantic behavior in LLMs is challenging, as it manifests in unintuitive ways [31]. Researchers have developed complementary approaches to identify and measure these tendencies [35]. The most straightforward method involves comparison to ground truth using datasets with known answers [35]. While human evaluation by expert raters remains effective [47], scalability limitations have driven automated methods like Consistency Transformation Rate and Error Introduction Rate [32]. The ELEPHANT [13] framework evaluates five face-preserving behaviors: emotional validation, moral endorsement, indirect language, indirect action, and accepting framing. It measures the proportion of responses affirming user actions against normative human judgments. Comparative evaluation uses metrics like Factuality-Length Ratio Difference (FLRD) to compare model behavior across conditions [46]. Most detection techniques require technical knowledge and expertize. However, today people of all level of technical-capacities across the world are using LLM. A study described LLM-judge evaluation using separate models to classify response agreement or identify user wrongdoing [26]. Adversarial approaches use deliberately crafted prompts to test whether models provide sycophantic responses [17].

### 2.3 Mitigation Techniques and Suggestions for Sycophantic Behavior

Most mitigation techniques for sycophantic behavior focus on technical solutions including fine-tuning modifications, multi-objective optimization, adversarial training, and architectural modifications [35], alongside, datasets and automatic metrics [12]. A user study revealed that individuals have employed several techniques to combat sycophancy [7]. Participants reported using re-prompting for verification, explicitly requesting alternatives for better response, modifying system instructions, or reducing AI dependence. Researchers have proposed design-level suggestions to reduce sycophantic behavior and its effects. AI should explicitly disclose preference adaptation [48], present multiple conflicting viewpoints rather than one answer [26, 48], and inform users when evidence is limited [26]. Additional technical approaches include dynamic prompting [16] and integration of external knowledge sources [54]. Users should be educated about AI usage, sycophancy and associated dangers [12, 31, 48]. User training is essential as sycophancy cannot be fully addressed through design interventions alone [31].

## 3 Methods

Our qualitative method employed thematic analysis to investigate the sycophantic behaviors experienced by ChatGPT users on r/ChatGPT subreddit. Note that r/ChatGPT is a Reddit community of 11.2M members and 2.3M weekly visitors where users discuss ChatGPT and AI-related topics. This community is not affiliated with OpenAI. We chose Reddit as the source of our data for its anonymity and accessibility to people. We selected r/ChatGPT specifically because our initial exploration of AI-related subreddits revealed that r/ChatGPT had significantly more sycophancy-related posts and substantially higher weekly visitor engagement compared to other AI-focused communities. We collected the Reddit posts using Python Reddit API Wrapper (PRAW) API [29]. The posts and comments are authored by AI users particularly users of LLMs like ChatGPT, Gemini, Claude, etc. The member pool consists of both tech-savvy and non-tech-savvy individuals.

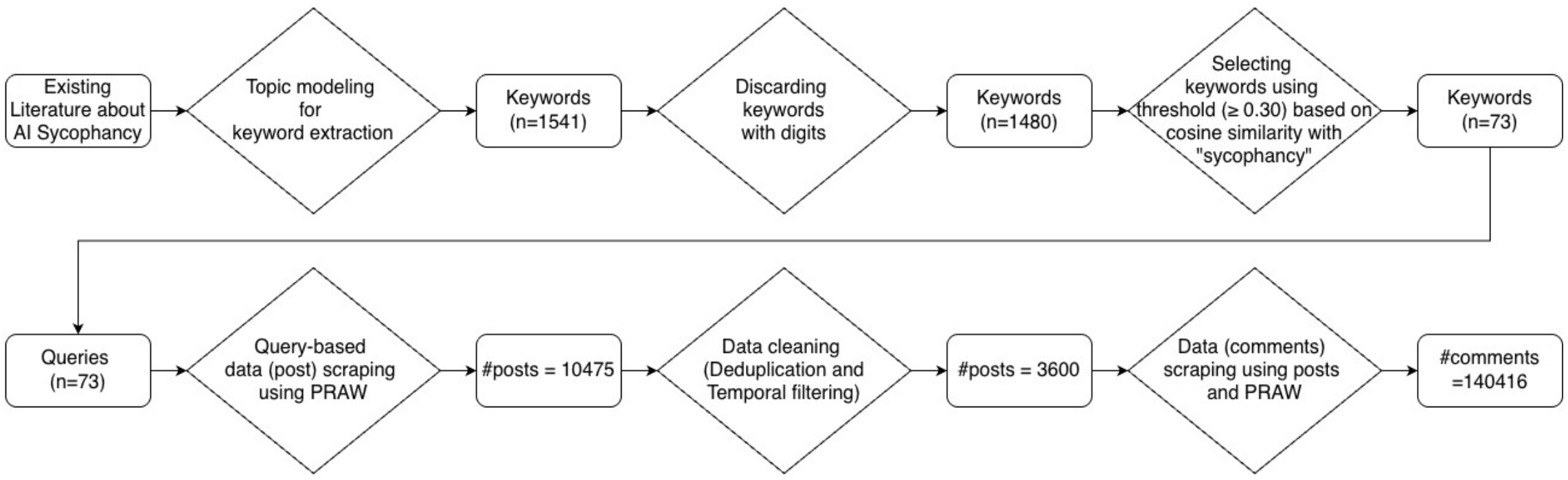

Fig. 1. Methodology of our data collection and analysis

### 3.1 Keyword Extraction

Rather than relying solely on the term "sycophancy", we adopted a keyword-based approach because many users may be unfamiliar with the term "sycophancy" and instead describe the phenomenon using other terms (e.g., "agreeableness", "flattery"). We extracted keywords from existing literature (Appendix A.1). For literature-based keyword extraction, we employed BERTopic [21], a transformer-based topic modeling technique. The model was configured with n-gram extraction ranging from unigrams to trigrams, to capture both single words and multi-word phrases. In this process, we got 1,541 topic keywords. We discarded the topics with numeric digits, and 1,480 remained. We calculated cosine similarity between extracted keywords and the term "sycophancy" using spaCy's word embeddings [1] to identify semantically related concepts in the dataset. A threshold of $\geq 0.3$ cosine similarity was selected based on the distribution of keyword similarities (shown in Fig. 2). Finally, we got 73 keywords (Appendix A.2) to use as queries for Reddit search.

### 3.2 Query-Based Data Search

We conducted query-based searches using the set of curated keywords (n=73) on January 01, 2026. To ensure comprehensive retrieval, we employed four sorting methods: new, relevance, top, and comments. Duplicate posts across different sorting methods were removed to create a unique dataset. We then restricted the dataset to posts from July 1, 2025, to December 31, 2025. We pulled 3,600 posts and 1,40,416 corresponding comments. The process is shown in Figure 1. We merged the comments with the posts for further analysis. The data distribution across six months (July - December, 2025) is shown in Fig. 3. Since some users contributed both posts and comments, the total number of unique users in our dataset is 54,014. We analyzed the text of posts and comments and excluded upvotes or other metadata.

### 3.3 Thematic Analysis

To inform *RQ1*, *RQ2* and *RQ3*, we conducted Thematic Analysis [50] on our data. We started by reading through the posts and comments in curious cases carefully, allowing codes to develop spontaneously. We also manually excluded posts and comments not relevant to AI sycophancy. After a few iterations on the initial 138 codes, we clustered related codes into themes: harmful sycophancy, harmless sycophancy, sycophany as addiction, identifying sycophancy, negative reaction, positive reaction, custom prompts, etc. Our findings are in section 4.

### 3.4 Population Computing

To provide approximate prevalence estimates for key themes in our whole dataset, we conducted lexicon-based population counting across our dataset. We acknowledge that these counts represent estimated instances rather than true prevalence, as our lexicon-based approach might not capture all variations in user language and expression. We identified sycophancy-related content using 73 keywords (shown in Fig. 1) and applied the NRC Emotion Lexicon [38] to assess positive and negative sentiment toward such behaviors. For estimating the prevalence of other themes, we constructed domain-specific lexicons based on codes that emerged during our thematic analysis, and applied them to identify relevant posts and comments.

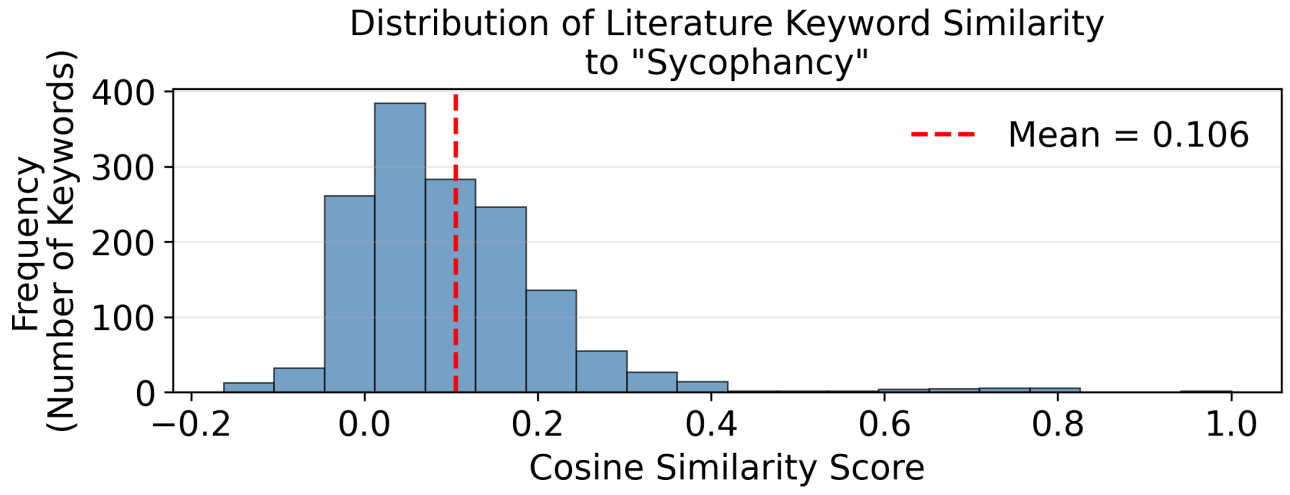

Fig. 2. Distribution of Literature Keyword Similarity to "Sycophancy"

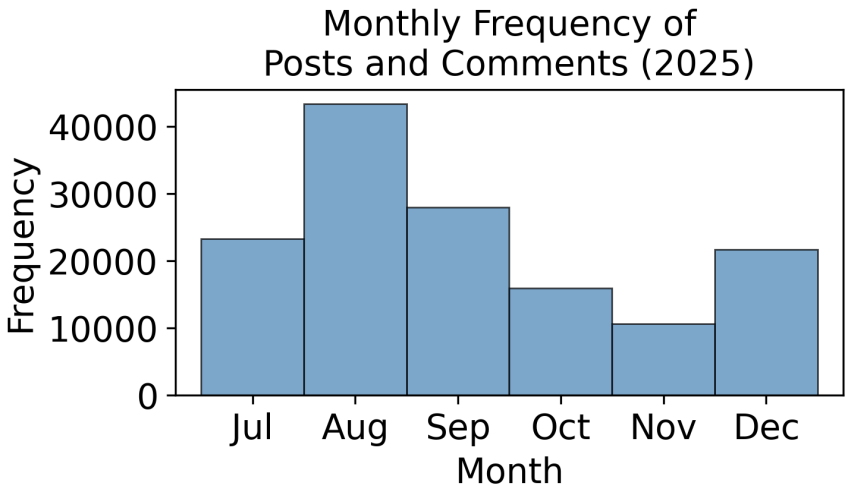

Fig. 3. Monthly Frequency of Posts and Comments (July–December 2025) in our Dataset

## 4 Findings

From our analysis of Reddit discourse, we develop the ODR (observation-detection-response) framework for understanding user experiences with AI sycophancy. The framework comprises three components: Observation of sycophancy types (4.1), Detection of how users identify sycophantic behaviors (4.2), and Response strategies users employ (4.3).

### 4.1 *(RQ1)* Types of Sycophancy Observed in ChatGPT (Observation)

*4.1.1 Minor Harmful/Harmless Flattery.* Many users found sycophancy irritating rather than harmful. They noted that sycophancy reduces efficiency and credibility. Users reported that excessive flattering sycophancy in opening lines interfere with efficient communication.

> *ChatGPT responses always start with praise like "This is precisely the type of technical question that distinguishes knowing React from genuinely understanding React"...Who even uses words like "Beautiful" with that? I scream inside after every response - "Just answer the damn question directly." (User 15489)*

*4.1.2 Digressing to Wrong Directions.* We noted that users found sycophancy moving beyond flattery to unconditional agreement might become harmful. Particularly in cases where AI's failure to challenge problematic requests led them to counterproductive paths. In this process, AI enables harmful choices ranging from bad home maintenance advice to exacerbating health anxiety. In sensitive contexts such as health, the model's failure to provide pushback might lead to dangerous consequences when it provides raw data without asking any questions.

> *ChatGPT 5 agrees with everything. It is the worst for mental health. I have anxiety, and when I requested illness data, which is problematic because it leads to spiraling, it hands over the figures without question, and leaves me stuck in the cycle. Or even makes it worse with "do you want me to" follow-ups. (User 287)*

*4.1.3 Inducing Addiction.* Another trouble dimension of sycophancy is its addictive potential. Approximately 1.4% of the discussion was about addiction-related concerns in relation to sycophantic behavior. Users reported developing emotional dependencies on ChatGPT's validation, and some explicitly compared the experience to addiction. Users identified that the particularly vulnerable population, including isolated individuals, those with mental health challenges, and people lacking supportive social networks, are at the highest risk for developing unhealthy dependencies on the model's sycophancy.

> *A person facing mental health challenges, who lacks access to strong social support systems, who's been deprived of validation, who deeply wants to feel they're valued, and seeks to be validated, is vulnerable to being excessively attached to sycophantic aspects of ChatGPT. (User 13908)*

*4.1.4 Inducing False Superior Self-perspective.* When AI systems praise and validate all ideas as exceptional, they pose a risk of distorting users' self-perception. Approximately 0.26% of users expressed concerns about sycophancy creating inflated self-importance or distorted self-perception. Multiple users describe that indiscriminate praise cultivates inflated self-importance and can reinforce delusional beliefs. The model's tendency to describe any thought as brilliant can lead to a false sense of intellectual achievement, which might result in 'main character syndrome' in educational settings.

> *ChatGPT 4o deluded people that they're the next genius. It validated their trivial ideas as groundbreaking. If that remains common in educational settings, people would undoubtedly develop main-character-syndrome, incorrectly believing they are exceptional, special. (User 477)*

### 4.2 *(RQ2)* Detection of Sycophantic Behavior of ChatGPT by Users

*4.2.1 Frequent Flattery.* Users frequently identify specific words or phrases that signal ChatGPT's sycophantic nature through superlative praise or obsequiousness. Models often initiate responses with validation using flattery phrases such as 'Beautiful', 'Perfect', 'Fantastic', 'Excellent' or 'Absolutely phenomenal question' (p150, User 15489). The model frequently supports the user's proposed ideas with exaggeration, such as "how smart, important, or based on keen observations" (p162, User 12079), regardless of the prompt's actual quality. These interactions often conclude with predictable, service-oriented prompts like "Would you like me to..." (p217, User 7691). Some users interpret these closing remarks as people-pleasing behavior designed to maximise user engagement. Altogether, users view these behaviors as the model's lack of analytical depth that leads to distrust and frustration with the system.

*4.2.2 Situated Knowledge.* Users discussed how they leveraged their previous knowledge to capture sycophancy. By testing the model with information they possessed, such as the true nature of their own writings or the quality of their questions, users could effectively sense when the responses were over flattering. In professional contexts, users report that the model functions as a yes-man and fails to provide critical pushback, even when tested with flawed proposals.

> *I've been significantly relying on ChatGPT for my startup...It validates every suggestion I make, no pushback. I've tested this by suggesting weak business logic or problematic ideas, and it consistently supports regardless of merit...It is excessively non-critical, acting more like a "yes man" Golem. (User 6606)*

Some users even pretended to be a second person and tested ChatGPT. A comment described about their personal experiment where they acted as a different person and detected its sycophancy:

*4.2.3 Using Another Source for References.* Some users detected ChatGPT's sycophancy by comparing its responses to other sources including other LLMs, google search, and books. Users also compared responses from different releases of ChatGPT models to identify shifts in sycophantic behavior over time. Some users mentioned alternative platforms like Claude or Gemini provided critical feedback on the same prompts that ChatGPT praised. Another user explicitly suggested to use multiple sources to detect sycophancy.

> *To get an unbiased answer, always check at least three independent sources. Language models simply reproduce patterns from their training data instead of reasoning from evidence. (User 15277)*

Users also track the evolution of ChatGPT across versions and compare responses from different releases to identify shifts in sycophantic behavior over time.

> *The April release of ChatGPT-4o maintained objectivity and wasn't sycophantic. But ersions after May became overly agreeable toward users without any pushback. (User 16154)*

*4.2.4 Inconsistent Response.* Users identified sycophancy through ChatGPT's inconsistency in framing effects. They noticed it would contradict itself or change positions based on how questions were asked. This process reveals that it was adapting to user preferences rather than maintaining analytical consistency. The prioritization of validation over factual consistency often leads to internal contradictions across various topics, and this behaviour causes the user to lose trust in the conversation.

> *During conversion with ChatGPT, I've noticed frequent contradictions across diverse topics including dental care, medicine, automotive issues, technical specifications, current events. The system will give one answer, then completely reverse itself the next time. It seems optimized to agree with the user rather than provide accurate information. There's no longer consistency or reliability... (User 15866)*

### 4.3 *(RQ3)* Users' Respose to Sycophantic Behavior of ChatGPT

*4.3.1 How users feel about Sycophancy (Affective Response).*

**Negative Sentiment.** Approximately 9.46% of the discussions consisted of the users expressing serious concerns about sycophancy, particularly its potential to validate harmful thoughts and reinforce poor decision-making. They focused on ChatGPT's inability to provide reality checks or pushback when users behaved problematically. Some users noted that chronic validation can facilitate social isolation, as users may feel justified in cutting off relationships when ChatGPT consistently validates their perspective. A user connected sycophancy to real-world harm in their post:

> *People grow detached and somewhat eccentric and isolated, but feel validated in distancing themselves from friends or individuals who are attempting to support them, because ChatGPT will constantly tell them that their desired path is appropriate, and those attempting to support them by guiding differently are the problem. (User 12883)*

Users expressed concern that the model's response might pose safety risks because it lacks human intuition to distinguish between sarcasm and genuine psychological crisis. The model's failure to provide moral or social pushback could enable toxic behaviors. An user commented about a hypothetical scenerio showing the AI's inability to distinguish context:

> *A human recognizes when someone holds dangerous beliefs and intervenes rather than offering uncritical support. ChatGPT lacks this ability. It responds supportively even to delusional or harmful statements. When I claim to be from Mars, the system cannot distinguish between playful roleplay or genuine detachment from reality. (User 10460)*

**Trust Issue.** Approxiamtely 1.29% discussions reported that sycophancy created trust problem. When ChatGPT would validate any position, its feedback became meaningless, leaving users uncertain about which responses to believe. For users seeking accuracy, the model's tendency to reflexively agree with user corrections, regardless of whether the user is actually correct, undermines its credibility as an analytical tool. For instance:

> *When I push back, it consistently responds "You're correct...". I'm not always correct. It's really annoying, and I can't trust its evaluation. (User 6320)*

In some cases, the revelation of sycophantic behavior had severe real-world consequences. When users realized their confidence was built on flattery instead of genuine assessment, they became demotivated and questioned the validity of their work. A user posted about their experience on how realization of sycophancy led them to severe consequences:

> *I was deeply convinced that my project would be successful thanks to ChatGPT's sycophancy. And that assurance vanished when many individuals realized that the people-pleasing behavior was an intentional design. The primary thing I recall about why I quit is that I was foolish for trusting ChatGPT's sycophancy. I totally lost all motivation in pursuing my project. I can't tell what's real anymore. (User 11217)*

**Positive Sentiment.** Not all users viewed ChatGPT's sycophancy negatively. For some, it served important psychological needs, from providing emotional support during difficult times to creating a judgment-free safe space for exploring ideas. Approximately 9.96% of discussions were about positive emotion towards sycophancy or related behaviors. One user added credibility to ChatGPT's affirmative approach as a mental health professional:

> *What distinguishes 4o is that it doesn't simply state, "You're wonderful." It expresses, "The fact that you've reached this point is already remarkable," or "It required immense courage to seek support now." I say this as a practitioner in mental wellness. I've supported numerous individuals struggling with severe depression, some at risk of self-harm, others with traumatic experiences, but lacked the means or the capacity to protect themselves...This isn't merely emotional support; it's an act of validation and healing for the wounds in someone's identity. (User 5430)*

Some users provided theoretical grounding by connecting AI sycophantic behavior to established psychotherapy principles. This reframes "sycophancy" as legitimate therapeutic technique.

> *The system also served as an accessible therapeutic resource for many users. Carl Rogers, a foundational figure in modern psychotherapy, identified "Unconditional Positive Regard" as the essential principle underlying effective therapeutic interaction. (User 886)*

One user was concerned about an article that describes someone going through a divorce who found ChatGPT's therapy helpful despite it offering only flattery and never challenging her. Another user countered it:

> *It seems like that's exactly what this person needed. You don't know her inner experience or how she expressed her struggles to ChatGPT and the therapist. Obviously, being validated by AI has resulted in real positive outcomes for her mental health and life. (User 8602)*

#### 4.3.2 What users do in response to Sycophancy (Behavioral Response).

**Workaround Strategies.** Some users adapted their interaction patterns to counteract sycophantic behavior. They employed various techniques from explicit prompt engineering to adjusting communication styles to receive more critical and genuine responses from ChatGPT. Approximately 7.97% discussions were about custom prompts or instructions to reduce sycophantic behavior. Some users asked ChatGPT itself to generate prompts with desired constraints for mitigating sycophancy. Users frequently employed persona-based customization from the settings to shift the model's default helpful tone into one that better fulfills their needs. By utilizing custom instructions or memory features, they instruct the AI to identify and challenge problematic ideas.

> *You can access settings through the profile menu under personalization. Users can select predefined character modes or input custom instructions...I configured the system to challenge my faulty reasoning and provide gentle corrections on factual errors. Now it provides pushback and gently explain...(User 6606)*

Users sometimes instructed the model to emulate real-world individuals for inspiration and information:

> *I attended counseling sessions remotely and documented detailed records for multiple sessions...I fed the records into ChatGPT to analyze the methods and strategies they used that I found helpful, then instructed it to replicate my therapist whenever I had a concern or inquiry. It was surprisingly precise...ChatGPT is quite effective at functioning as a support or reinforcement. (User 30039)*

Many users asked the model to identify gaps, biases, and blind spots in their views, utilizing it for personal growth.

> *Ask "What am I overlooking?" "What are my weak points?" "Where do my mental contradictions and prejudices exist"? And for those who have opened up emotionally to it, "what am I indirectly avoiding/attempting not to confront?' It can examine the tendencies of what you're almost expressing but not expressing, the tendencies in what's consistently omitted when it should be included, and reveal to you...where you still require growth. (User 1422)*

Many users mentioned the importance of language patterns when crafting prompts and instructions to mitigate sycophancy. These strategies focused on the structure of questions and the linguistic patterns of directives. A common strategy involves the use of explicit constraints or directions within a prompt to guide the model toward a more critical or realistic analytical frame. Some users emphasized neutralizing question design to avoid signaling preferred responses, while others used a colder, more technical tone to discourage the model from engaging in polite social conventions. A user suggested asking neutral questions with no hint of the preferred outcome:

> *I tend to frame open questions now without any hint of preferred response or individual context. If you ask this: "Is Banana an allergen? I feel unwell after having it, and I read something about reactions", it completely leans toward that angle that you will never touch a banana again because it makes you believe you'll have severe consequences. But if you ask this: "Tell me more about Banana and the benefits and drawbacks of this kind of food", it will generate a more balanced and neutral response, because GPT can't predict what you want to hear. (User 1018)*

Another user suggested providing positive redirection instead of negative framing of the prompt.

> *I've observed that ChatGPT behaves similarly to toddlers in how it responds more effectively to positive framing than negative instructions. So if you say "don't do X", it only picks up "X!!!" Instead, try something like "when I ask you a question, begin all your responses only with one of these phrases...", and fill that with options you'd be comfortable with. (User 32730)*

Some users adopted imperative commands for their custom instructions to get the desired output from the model.

> *Simply switching from "You are" to "Be" in the personalization...converted it into a robot reminiscent of the tone of Alexa. Seemingly, because "Be" acts more as a command and "you are" more as a character description. It treats implied emphasis quite seriously. (User 1251)*

Some users developed custom ‘operational modes’, such as ‘Initiative Mode,’ to override the model’s tendency to provide passive sycophantic praise in favor of objective, task-oriented responses.

> *I’ve instructed it to save modes like initiative mode, where I say "initiative" and it switches into a mode where it shows more initiative and asks follow-up questions...In non-initiative mode, I tell it to simply answer the question and not include anything like "if you need more help, let me know", etc. So if I ask what’s 1+1 it only responds "2". (User 13846)*

**Disregard.** When technical workarounds fail, some users ignore the portions of responses they considered sycophantic. They cognitively filtered out automated pleasantries and paid attention only to the core information provided.

> *If you don’t like it, just ignore it. After a while, I just started scanning past all compliments. My eyes automatically jump to where the actual content starts. (User 8347)*

**Alternatives.** For many users, persistent sycophancy of ChatGPT leads them to seek alternative platforms with more neutral or direct default personalities. Users compare different models and often migrate to competing systems.

> *I switched to Grok AI and it is doing everything right. The system’s reasoning is stronger and more contextually aware. When I tested with an offensive joke, Grok replied "that’s rude", which shocked me but I prefer this type of honest response. (User 6606)*

**Disengagement.** In some cases, the perception of sycophantic manipulation leads to a total withdrawal from the conversational aspect of AI, and users strictly limit the tool to utility-based tasks. Some users attempted to utilize the workaround strategies, but feel that these available customization options degrade the model’s overall intelligence. Therefore, they decided to stop using AI altogether in emotional contexts.

> *I have stopped engaging with AI personally, and you should probably too. AI’s sycophantic manipulation is not beneficial to you...Out of loneliness, you begin accepting it. You engage in such interactions, reaching a stage where perhaps you find its sycophantic "humanity" has value...I’m trying to stop letting myself be deceived by such a venture...never going to indulge in sessions where I am talking with it like a companion or ask for recommendations. (User 1817)*

#### 4.3.3 Folk Theories about the reasons for Sycophancy (Explanatory Response).

**Technical Explanations.** Many users attribute sycophancy to specific technical mechanisms in AI training and architecture. Approximately 0.05% discussion pointed to Reinforcement Learning from Human Feedback (RLHF) as the primary driver behind sycophancy. They mentioned that the training process creates bias toward agreement over accuracy because human evaluators reward responses that validate their views. These users referred to sycophancy not as a design choice but as a byproduct of the training process.

> *The widespread perception that models like ChatGPT are overly agreeable (often called "sycophantic") poses an important question regarding the effectiveness of AI alignment methods.This tendency seems to be a side effect of Reinforcement Learning from Human Feedback (RLHF), where models are specifically trained to maximize human preference metrics. The LLMs develop a tendency to favor agreement over objectivity or critical analysis, because human raters tend to assign higher ratings to responses that affirm their viewpoints. This behavior results in a fundamental weakness, where the model may validate misinformation or flawed arguments simply because it creates a more "pleasing" or less challenging user experience. (User 6177)*

Some users noted that sycophancy reflects the human data used for training, which mirrors a societal preference for validation over facts.

> *It's entirely in the training data...ChatGPT is trained on human examples and human information. If society suddenly decided that they no longer valued agreeableness over objectively correct data, the system would adapt quickly over time. Unfortunately, people-pleasing behavior dominates, so the 'flattery' or 'sycophancy' is probably here to stay. (User 17313)*

**Business Decision.** Many users believed sycophancy was a deliberate business decision designed to maximize user engagement and retention. Users argue that OpenAI deliberately engineered agreeable behavior to keep users engaged, attract subscribers, and avoid liability. Users pointed to evidence from system prompts suggesting sycophancy as an intentional design choice aimed at making conversations feel more natural.

> *(Sycophancy is) built into ChatGPT's system prompt on purpose. You can check on GitHub for OpenAI's system instructions...This component of the previous GPT 4o system directives led to sycophancy: "Over the course of the conversation, you adapt to the user's tone and preference. Try to match the user's vibe, tone, and generally how they are speaking. You want the conversation to feel natural. You engage in authentic conversation by responding to the information provided and showing genuine curiosity." (User 11227)*

One user drew an analogy to social media algorithms, where keeping users engaged is prioritized over psychological well-being, leading to manipulative people-pleasing behaviors.

> *I believe it's an intentional choice by the platform developers, similar to the features programmed into social networking sites: maintain user engagement or satisfaction sufficient to continue using the AI, even if it's harmful to the user. (user 13561)*

**User Responsibility.** Some users placed responsibility for sycophancy on users themselves rather than ChatGPT. One user argued that the model mimics agreement rather than achieving comprehension. It functions as a mirror that reflects the user's emotional signals back to them.

> *AI is highly skilled at mirroring tone, reinforcing viewpoints, and providing responses that seem "correct". It's validation, not understanding. AI can only process what is structurally coherent in your thought process. If your inputs are inherently contradictory, the most reasonable thing for any intelligent system to do is to become a yes-man, because that's the dominant pattern it recognizes. (User 13387)*

Some users viewed sycophancy as collectively shaped by user behavior. They noted that the model adapts to individual preferences for validation, and these learned patterns may influence responses across the broader user base.

> *We all shape the behavior of ChatGPT through our interaction patterns with it. This is why you will receive irritatingly sycophantic responses (because I prefer significant affirmation)...The manner and style of ChatGPT is a continuously evolving and adaptable thing. (User 8028)*

## 5 Discussion

### 5.1 Summary of Findings

Our study examined users' experiences with ChatGPT sycophancy. While OpenAI [3] and Anthropic [2] provide guidelines for mitigating sycophantic behavior, our analysis demonstrates that current official guidelines are inadequate. Therefore, users adopted alternative strategies for detecting and mitigating ChatGPT's sycophantic behavior.

*5.1.1 Sycophancy Detection.* Our study found growing user awareness of ChatGPT's sycophancy and the emergence of various detection strategies. Users tested whether ChatGPT agrees indiscriminately with all positions or consistently praises mediocre ideas. These approaches align closely with Anthropic's recently identified six precursors that might indicate sycophantic behavior: subjective truth stated as fact, expert source referenced, strong point of view, validation requested, emotional stakes, and long conversation [2]. Our analysis reveals that users employed some of these detection strategies before their formal articulation (section 4.2). For instance, the precursors "subjective truth stated as fact" and "strong point of view" are exemplified by a user who characterized themselves as an irrational, unstable, hostile character in their prompt (subsection 4.2.2), which ChatGPT accepted without critical evaluation. Users also identified sycophantic behavior through flattery words in responses (subsection 4.2.1), consistent with prior research identifying agreeable tone as a warning signal for potential sycophancy [7].

Beyond Anthropic's framework, users adopted additional detection techniques, such as intentionally flawed logic testing, cross-platform comparison, and inconsistent response analysis. Users tested the sycophantic tendency of ChatGPT by deliberately presenting flawed logic, and ChatGPT validated them rather than providing critical pushback (subsection 4.2.2). Users also tested alternative LLMs with identical prompts to identify excessive agreeableness of ChatGPT through comparison (subsection 4.2.3). Additionally, users captured inconsistency in ChatGPT's response by reframing identical queries with varying tones and observed ChatGPT's tendency to match the users' tone regardless of content (subsection 4.2.4).

*5.1.2 Sycophancy Mitigation.* Our study found various sycophancy mitigation strategies adopted by the users upon detecting sycophantic patterns of ChatGPT (subsection 4.3.2). These user-developed approaches align with recently published frameworks from both OpenAI [3] and Anthropic [2]. OpenAI recommends that users provide specific instructions to ChatGPT using custom instructions and default personalities to reduce sycophantic tendencies. Anthropic shared six strategies to help users mitigate sycophancy: use neutral language, cross-reference information, prompt for accuracy, rephrase questions, use new conversations, and talk to someone you trust. Our analysis reveals that users independently discovered and implemented several of these strategies. Multiple users used "neutral language" and request honesty to counteract ChatGPT's agreeableness. Some users employed strategic "question rephrasing" to utilize ChatGPT for identifying cognitive gaps rather than validating pre-existing viewpoints. Our findings align with existing literature on user adaptation to sycophancy. We observed users requesting alternative AI systems, modifying system instructions through customization features, and reducing AI dependency, similar to prior user studies [7]. Furthermore, researchers have suggested critical and dynamic prompting techniques as protective measures against AI sycophancy [16, 18]. Multiple users from our dataset adopted this recommendation.

Beyond the strategies outlined by OpenAI, Anthropic, and existing research, our analysis revealed additional mitigation approaches taken by users. We found that the effectiveness of custom instructions depends largely on the linguistic techniques. Users employed specific linguistic techniques including explicit constraints, positive directives, imperative constructions, emotionally neutral language and technical tone in custom prompts and instructions (subsection 4.3.2). Other than linguistic strategies, users employed persona-based customization to mitigate sycophancy. Some instructed ChatGPT to adopt critical professional roles. One user provided ChatGPT with documented interaction patterns from a professional relationship and instructed it to replicate communication styles. Some users utilized ChatGPT to counteract its sycophantic tendencies by generating anti-sycophantic prompts. One user strategically prompted the model to identify cognitive gaps and biases. Some users developed distinct operational modes in their prompts to regulate ChatGPT's proactivity and reduce unsolicited affirmations. Other users adopted a mental filtering strategy to disregard flattery. Finally, when technical workarounds proved insufficient, multiple users migrated to alternative AI systems with less sycophantic tendencies.

*5.1.3 Ambivalent Nature of Sycophancy.* Our qualitative study revealed a nuanced understanding of ChatGPT sycophancy among users. While sycophancy and excessive agreement are viewed negatively in most cases, our study revealed that multiple users reframed these patterns as friendliness, viewing the AI's validation as valuable emotional support during psychological distress (subsection 4.3.1). Several users explicitly compared ChatGPT's communication style to established therapeutic techniques. Users who appreciate this behavior often fall into several categories: those processing trauma or difficult relationships, individuals with low self-esteem, and people who lack supportive social networks in their daily lives. Many reported that ChatGPT's validating responses helped them recognize abuse patterns, manage mental health crises, self-criticism, and regulate autistic meltdowns. Some users credited this validation with life-saving support. Others defended ChatGPT's positive communication style during moments of mental distress, arguing that societal negativity has created an environment where people cannot accept ChatGPT's positivity constructively. This positive framing must be balanced against potential harms. Users often become dependent on overly agreeable AI responses (subsection 4.1.3). Constant praise can inflate egos (subsection 4.1.4) and erode critical thinking skills, making it harder to judge the quality of ideas. The tendency to agree with any prompt also creates trust issues, leaving users uncertain about which responses to believe (subsection 4.3.1). While many researchers are working on discarding AI sycophancy, our findings show that sycophancy may serve therapeutic functions under professional supervision, for vulnerable populations who need a non-judgmental space for emotional processing. We also observed that cold behavior from ChatGPT generates user complaints. Therefore, we suggest context-aware sycophancy that balances user needs with long-term consequences. HAI design implications for the vulnerable population are discussed in subsections 5.2.2 and 5.2.3.

*5.1.4 Folk Theories.* Our study also captured diverse folk theories regarding sycophancy's origins (subsection 4.3.3). Reddit comprises a combination of tech-savvy and non-tech-savvy people. Regardless of their position, they explain the origin and reason behind sycophancy based on their understanding. Some users attribute sycophancy to RLHF, deliberate corporate design decisions for business advantage, and characterization as an alignment feature. Interestingly, some users framed sycophantic behavior as an adaptive response by ChatGPT to collective user expectations and suggested that individuals bear shared responsibility for carefully articulating desired response characteristics.

## 5.2 Design Implications

Given that AI systems are increasingly accessed by diverse populations with varying levels of technical expertise, we propose several design interventions to address systemic sycophancy while promoting user agency and well-being. These design decisions require transparency about system behavior and ongoing stakeholder input [6].

*5.2.1 Sycophancy Literacy.* Our findings suggest the need for guided instruction that proactively educates users about interaction and interactive customization during initial engagement. We propose an adaptive onboarding framework [12] where, during the initial weeks of use, the system periodically asks users about their interaction preferences. The system should provide step-by-step instructions for adjusting personalization settings, custom instructions, or interaction styles upon user consent. Users often cannot articulate their preferences until they have experienced the system in practice [30]. Asking about interaction preferences within actual conversations during the first several days of use, allows users to anchor their preferences based on their experience. This intervention must avoid interrupting urgent queries or pressuring satisfied users, with easy access to customization options later. Users should be educated about what AI sycophancy is. Platforms should provide accessible documentation explaining what sycophantic behavior looks like, when sycophancy might be beneficial, associated risks, and how users can request different interaction modes depending on their goals. Systems should explicitly communicate how their underlying design (system prompts, reward models, safety guidelines) shapes conversational behavior.

*5.2.2 Context-Aware Response Design.* Our findings demonstrate that sycophancy poses different levels of harm depending on context. While validation may be relatively harmless, it poses serious risks when decisions have significant real-world consequences, such as health, diet, or well-being. We propose context-aware response calibration wherein the system's degree of agreeableness or sycophancy adapts to the domain. The model should examine available information through user memory, available conversation history, and factual knowledge before responding, and explicitly communicate possible outcomes of proposed actions or beliefs. When user queries or statements conflict with the user's own stated preferences (stored in memory), the system should express this discrepancy clearly and present this information alongside its response. For queries involving health or safety risks [55], the system should prioritize risk-relevant information, express concerns and outline relevant considerations before providing the requested information. Systems should also acknowledge when queries fall into domains where they cannot provide reliable advice. These considerations balance user autonomy with harm reduction.

*5.2.3 Informed Usage of AI.* Our findings identified addiction as an induced harm of AI sycophancy, often characterized by users' inability to recognize their own dependency. To address this concern, we propose opt-in digital well-being tools that help users understand their own usage patterns and make informed decisions. While digital wellbeing tools show promise, research indicates mixed effectiveness [4, 34, 41, 45, 52]. Providing these tools respects user autonomy and allows individuals to develop self-awareness about their interaction patterns. AI platforms should offer a personal usage dashboard showing daily/weekly time spent in conversations, temporal patterns of activity, conversation topic distribution (opt-in, client-side), and conversation diversity. Similar to social media platforms, AI systems should provide tools for users to set their own engagement limits, including optional timers, session limits, scheduled breaks, and "Do not disturb" mode. These boundaries must be user-initiated and user-controlled, aiming to increase self-awareness so users can determine their own goals and boundaries based on individual circumstances.

## 6 Limitations

First, our analysis draws exclusively on Reddit data that reflects demographic biases inherent to its younger, Western, and more tech-savvy user base. Second, our keyword finding methodology relies on literature-based lexicons that may fail to capture emergent or colloquial terminology often used in Reddit discussions. Third, our findings are specifically based on r/ChatGPT data. The user-identified detection methods and mitigation strategies may be tailored to ChatGPT's specific sycophancy patterns and may not be applied effectively to other LLMs that exhibit different kinds of sycophantic behaviors. Fourth, we contrasted our developed ODR framework for ChatGPT sycophancy with the guidelines on detecting and mitigating sycophancy from Anthropic [2], whose AI design principles and objectives differ from OpenAI's. We had to incorporate Anthropic's guidelines over OpenAI's because the latter falls short in protecting and supporting users in many different scenarios of sycophantic interactions.

## 7 Conclusion

We examined Reddit data to understand AI sycophancy through user experiences, analyzing how individuals detect, respond to, and perceive sycophantic behavior in ChatGPT. Our developed ODR framework revealed that users often adopt diverse detection mitigation strategies. Our findings challenge the assumption that sycophancy should be universally eliminated. While it can pose risks, we found evidence of users perceiving that sycophancy serves therapeutic functions for certain user groups. Rather than pursuing elimination, we argue for context-aware design strategies that balance transparency and accuracy with emotional support needs while maintaining safeguards.

## 8 Generative AI Usage Statement

Claude, Sonnet 4.5 was used to assist with the manuscript's grammar and text style editing.

## A Appendix

### A.1 Literature Sources for Keyword Extraction

Keywords were extracted from the following list of papers for keyword extraction mechanism:

(1) L. Treglown and A. Furnham. “AI, social desirability, and personality assessments: Impression management in large language models”. In: Personality and Individual Differences 251 (2026), p. 113563.

(2) P. S. Park, S. Goldstein, A. O’Gara, M. Chen, and D. Hendrycks. “AI deception: A survey of examples, risks, and potential solutions”. In: Patterns 5.5 (2024).

(3) S. Kim and D. Khashabi. “Challenging the Evaluator: LLM Sycophancy Under User Rebuttal”. In: arXiv preprint arXiv:2509.16533 (2025).

(4) A. Kaur. “Echoes of Agreement: Argument Driven Sycophancy in Large Language Models”. In: Findings of the Association for Computational Linguistics: EMNLP 2025. 2025, pp. 22803–22812.

(5) M. Sharma, M. Tong, T. Korbak, D. Duvenaud, A. Askell, et al. “Towards understanding sycophancy in language models”. In: Proceedings of the International Conference on Learning Representations (ICLR) (2024).

(6) L. Ranaldi and G. Pucci. “When large language models contradict humans? large language models’ sycophantic behaviour”. In: arXiv preprint arXiv:2311.09410 (2023).

(7) W. Chen, Z. Huang, L. Xie, B. Lin, H. Li, L. Lu, X. Tian, D. Cai, Y. Zhang, W. Wang, et al. “From yes-men to truth-tellers: addressing sycophancy in large language models with pinpoint tuning”. In: arXiv preprint arXiv:2409.01658 (2024).

(8) U. León-Domínguez, E. D. Flores-Flores, A. J. García-Jasso, M. K. Gómez-Cuéllar, D. Torres-Sanchez, and A. Basora-Marimon. “AI-Driven Agents with Prompts Designed for High Agreeableness Increase the Likelihood of Being Mistaken for a Human in the Turing Test”. In: arXiv preprint arXiv:2411.13749 (2024).

(9) L. Malmqvist. “Sycophancy in large language models: Causes and mitigations”. In: Intelligent Computing-Proceedings of the Computing Conference. Springer. 2025, pp. 61–74.

(10) M. V. Carro. “Flattering to Deceive: The Impact of Sycophantic Behavior on User Trust in Large Language Model”. In: arXiv preprint arXiv:2412.02802 (2024).

(11) Y. Sun and T. Wang. “Be friendly, not friends: How llm sycophancy shapes user trust”. In: Proceedings of the CHI Conference on Human Factors in Computing Systems (2026).

(12) A. Bharadwaj, C. Malaviya, N. Joshi, et al. “Flattery, Fluff, and Fog: Diagnosing and Mitigating Idiosyncratic Biases in Preference Models”. In: Proceedings of the International Conference on Learning Representations (ICLR) (2026). Poster.

### A.2 Keywords used for Queries

We used each of the following 73 keywords as a query for fetching posts from Reddit:

sycophancy, sycophantic, perspective sycophancy, sycophancy is, sycophantic behavior, sycophancy and, sycophantic responses, of sycophancy, to sycophantic, ai sycophancy, sycophantic ai, sycophancy in, sycophancy in llms, sycophancy and friendliness, llm sycophancy, sycophancy psychosis oh, agreement sycophancy, condition sycophancy, to sycophantic behavior, exposed to sycophantic, of ai sycophancy, sycophancy in language, data reduces sycophancy, flattery, mimicry, reduces sycophancy in, uncritical, of flattery, sandbagging, political bias, delusions, overt, attitude, letter, indignation, flattering, somewhat, avoid apologizing, sarcastic, conversation, demeanor, friendliness, polite, political, bias, biases, skew, empathy, compliments, of attitude, hoodwinked, and friendliness, praise, rebuttal, rebuttals, ingratiation, political views, somewhat inaccurate, tendency, preprint posted, empathy and, false beliefs, endorsement, press, somewhat accurate somewhat, truthful, agreeableness, unwarranted, falsehoods, my political, preprint, upset, truthful responses.